\documentclass[superscriptaddress,twocolumn,prl,amsmath,amssymb]{revtex4}
  \usepackage{pgfplots} 
  \usepackage{ulem}
  \usepackage{color}
  \usepackage{amssymb}
  \usepackage{amsmath}
  \usepackage{amsthm}
  \usepackage{physics}
  \usepackage{bbold}
  \usepackage{bm}
  \usepackage{verbatim}
  \usepackage[utf8]{inputenc}
  \usepackage{bbm}
  \usepackage{svg}
  \usepackage{soul}

  \usepackage{gensymb}
  \usepackage{float}
  \usepackage{hyperref}
\newcommand{\be}{\begin{eqnarray}}
\newcommand{\ee}{\end{eqnarray}}

\definecolor{ao(english)}{rgb}{0.0, 0.5, 0.0}
\definecolor{awesome}{rgb}{0.85, 0.13, 0.12}
\definecolor{amber}{rgb}{1.0, 0.75, 0.0}
\definecolor{babyblue}{rgb}{0.54, 0.81, 0.94}
\definecolor{babypink}{rgb}{0.96, 0.76, 0.76}
\definecolor{bluebell}{rgb}{0.64, 0.64, 0.82}
\definecolor{burgundy}{rgb}{0.5, 0.0, 0.13}
\definecolor{green}{rgb}{0.1, 0.45, 0.036}
\definecolor{amber(sae/ece)}{rgb}{1.0, 0.49, 0.0}
\pgfplotsset{compat=1.11,
        /pgfplots/ybar legend/.style={
        /pgfplots/legend image code/.code={%
        \draw[##1,/tikz/.cd,bar width=3pt,yshift=-0.2em,bar shift=0pt]
                plot coordinates {(0cm,0.8em)};},
},
}

\begin{document}

\title{Correcting for finite statistics effects in a quantum steering experiment}
\author{Sophie Engineer}
\affiliation{Quantum Engineering Centre for Doctoral Training, H. H. Wills Physics Laboratory and Department of Electrical \& Electronic Engineering, University of Bristol, Bristol, United Kingdom}
\affiliation{Institute of Photonics and Quantum Sciences (IPAQS), Heriot-Watt University, Edinburgh, United Kingdom}

\author{Ana C. S. Costa}
\affiliation{ Department of Physics, Federal University of Paran\'a,  Curitiba, PR, Brazil}
\affiliation{ Department of Exact Sciences and Technology, State University of Santa Cruz, Ilhéus, BA, Brazil}

\author{Alexandre C. Orthey Jr.}
\affiliation{ Department of Physics, Federal University of Paran\'a,  Curitiba, PR, Brazil}
\affiliation{ Center for Theoretical Physics, Polish Academy of Sciences, Warsaw, Poland}
\affiliation{Institute of Fundamental Technological Research, Polish Academy of Sciences, Pawi\'nskiego 5B, 02-106 Warsaw, Poland}

\author{Xiaogang Qiang}
\affiliation{National Innovation Institute of Defense Technology, AMS, Beijing, China}

\author{Jianwei Wang}
\affiliation{State Key Laboratory for Mesoscopic Physics, School of Physics, Peking University, Beijing, China}
\affiliation{Frontiers Science Center for Nano-optoelectronics \& Collaborative Innovation Center of Quantum Matter, Peking University, Beijing, China}

\author{Jeremy L. O'Brien}
\affiliation{Department of Physics, The University of Western Australia, Perth, Western Australia, Australia}

\author{Jonathan C.F. Matthews}
\affiliation{Quantum Engineering Technology Labs, H. H. Wills Physics
Laboratory and Department of Electrical \& Electronic Engineering,
University of Bristol, Bristol, United Kingdom}

\author{Will McCutcheon}
\affiliation{Institute of Photonics and Quantum Sciences (IPAQS), Heriot-Watt University, Edinburgh, United Kingdom}

\author{Roope Uola}
\affiliation{D\'epartement de Physique Appliqu\'ee, Universit\'e de Gen\`eve, Gen\`eve, Switzerland}

\author{Sabine Wollmann}
\affiliation{Institute of Photonics and Quantum Sciences (IPAQS), Heriot-Watt University, Edinburgh, United Kingdom}
\affiliation{Quantum Engineering Technology Labs, H. H. Wills Physics
Laboratory and Department of Electrical \& Electronic Engineering,
University of Bristol, Bristol, United Kingdom}

\begin{abstract}

Verifying entanglement between parties is essential for creating secure quantum communication. However, finite statistics can lead to false positive outcomes in any tests for entanglement. Here, we introduce a \textcolor{black}{one-sided} device-independent protocol that corrects for apparent signaling effects in experimental probability distributions, caused by statistical fluctuations and experimental imperfections. We use semidefinite programming to identify the optimal inequality, for our experimental probability distribution, without resource-intensive tomography. Our protocol is numerically and experimentally analysed in the context of random, misaligned measurements, correcting apparent signaling where necessary. Our results show a significantly higher probability of violation than existing state-of-the-art inequalities. This work demonstrates the power of semidefinite programming for entanglement verification and brings quantum networks closer to practical applications.

\end{abstract}

\maketitle



{\it Introduction.---} Quantum networks have applications across many sectors, from secure communications \cite{wei2022towards, mehic2020quantum} to distributed sensing \cite{guo2020distributed} and the development of a quantum internet \cite{Kimble2008, simon2017towards}. Thanks to recent technological advances, such as ultra-bright photon sources and sophisticated detectors, quantum information processing tasks in quantum networks are now becoming a reality~\cite{Kimble2008, Wehner2018, Tavakoli2021}. This has been successfully demonstrated in free-space and fibre-base quantum networks~\cite{Yin2020, boaron2018secure}.
The most rigorous way to verify non-locality in networks are Bell tests, which are fully-device independent and treat measurement devices as black boxes~\cite{hensen2015,giustina2015,shalm2015,rosenfeld2017}. However, despite today's technological advancements, Bell tests remain demanding for realistic quantum networks.

Thus, quantum steering -- a relaxation of Bell tests, has received a substantial amount of attention~\cite{Wiseman2007, Uola2019, Cavalcanti2016review} thanks to its noise tolerance and loss robustness~\cite{srivastav2022noise, Weston2018}. Steering based tasks are \textcolor{black}{one-sided} device independent (\textcolor{black}{1s}-DI), meaning that one party’s measurement devices are trusted, while the other party’s devices are untrusted and treated as a black box. This allows the trusted party, Bob, to perform quantum state tomography and gain full information about his state, while the untrusted party, Alice, is limited to her conditional measurement outcomes. 

Experimental demonstrations of such nonlocality tests assume that the collected statistics are not biased to any experimental imperfections and represent a fair sample of the true underlying distribution of the shared quantum states. Additionally, reduced transmission rates in long-distance quantum communication, in combination with imperfect detector technology, can result in scenarios where only limited data can be collected, increasing the effect of finite statistics.
For practical applications of nonlocality in quantum networks, these limitations need to be reduced. This either requires careful experimental designs \cite{nadlinger2022experimental, zhang2022device} and the use of advanced detection technology, or demands more sophisticated numerical approaches.

Here, we develop a resource-efficient protocol that allows us to demonstrate steering 
by numerically correcting for finite statistics effects. 
Our protocol is based on the steering robustness (SR)~\cite{Piani2015}, a quantifier of quantum steering that uses semidefinite programming (SDP) and resource-intensive full state tomography on Bob's side. Using this measure rather than the binary test of steering inequalities informs us about the noise robustness of an entangled state~\cite{skrzypczyk2014quantifying}.
In comparison to the SR, our protocol -- the adapted steering robustness (ASR) -- is resource-minimal, as it does not require tomography on either party. Furthermore, it numerically searches for the tightest, optimal, steering inequality for the recorded data and is efficiently solvable on a classical computer. We find, in simulations, that for a maximally entangled state, the ASR outperforms other compared steering inequalities when Bob and Alice perform two misaligned orthogonal measurements each. For the three settings per party case, the ASR is optimal.

We experimentally test our protocol on a robust photonic quantum processor~\cite{Qiang2018} with the stability to perform 10,000 random orthogonal measurements per side. Our method requires the shared quantum correlations to be non-signaling~\cite{popescu1994, Horodecki2019, Gisin2020}, which is a built-in \textcolor{black}{constraint} for any steering criterion. Without satisfying this \textcolor{black}{constraint}, there is the potential for false positives to occur and this must be carefully considered. Due to finite statistics, experimental demonstrations rarely fully satisfy this principle~\cite{renou2017inequivalence, smania2018avoiding} and our analysis involves a numerical optimisation method. Our \textcolor{black}{1s}-DI non-signaling model algorithm (NSMA) finds the closest, according to a chosen order of merit, non-signaling statistics, that we assume to be the underlying behaviour of our experiment. After applying the NSMA to our experimental statistics, we compare the ASR to other well-established quantum steering inequalities~\cite{Cavalcanti2009,Saunders2010, Moroder2016, Wollmann2018} and find that the ASR performs significantly better than all other inequalities.

{\it Results.---} 
In a steering scenario, one assumes that two parties, Alice and Bob, share a quantum state $\rho_{AB}$. Bob tries to establish if the state is entangled, without trusting Alice's measurement outcomes. In each iteration of the protocol, Bob receives his part of the shared state and announces randomly chosen measurement settings $x \in \{1,...n\}$ for Alice to perform locally on her part of the state. Then, Alice declares her corresponding measurement outcome $a$ to Bob. Finally, Bob has access to his collection of post-measurement states $\{\varrho_{a|x}\}_{a,x}$ along with Alice's conditional probabilities $\{\operatorname{p}(a|x)\}_{a,x}$. This information is expressed in the set of states, known as the state assemblage $\sigma_{a|x}=\operatorname{p}(a|x)\varrho_{a|x}$. Alice's measurements are described by a positive-operator valued measure (POVM) $\{A_{a|x}\}_{a}$ for each $x$, i.e. $\sum_a A_{a|x}=\openone$ and $A_{a|x}\geq0$ for each $a$, where $\openone$ is the identity, and $\rho_{AB}$ is the state shared between Alice and Bob. Given this formalism, members of the assemblage set can be described by $\sigma_{a|x}:=\text{tr}_A[\left(A_{a|x}\otimes\openone \right)\rho_{AB}]$. Bob can use the assemblage to test if Alice's results can be explained by a local hidden state (LHS) model~\cite{Wiseman2007}.
If such an LHS model does not exist, the shared state $\rho_{AB}$ is entangled. Thus, Alice can steer his system via her measurements. In this scenario, Bob is required to have complete knowledge of his post-measurement state $\varrho_{a|x}$.
\begin{figure}[htbp!]
    \centering
\includegraphics[width=0.36\textwidth]{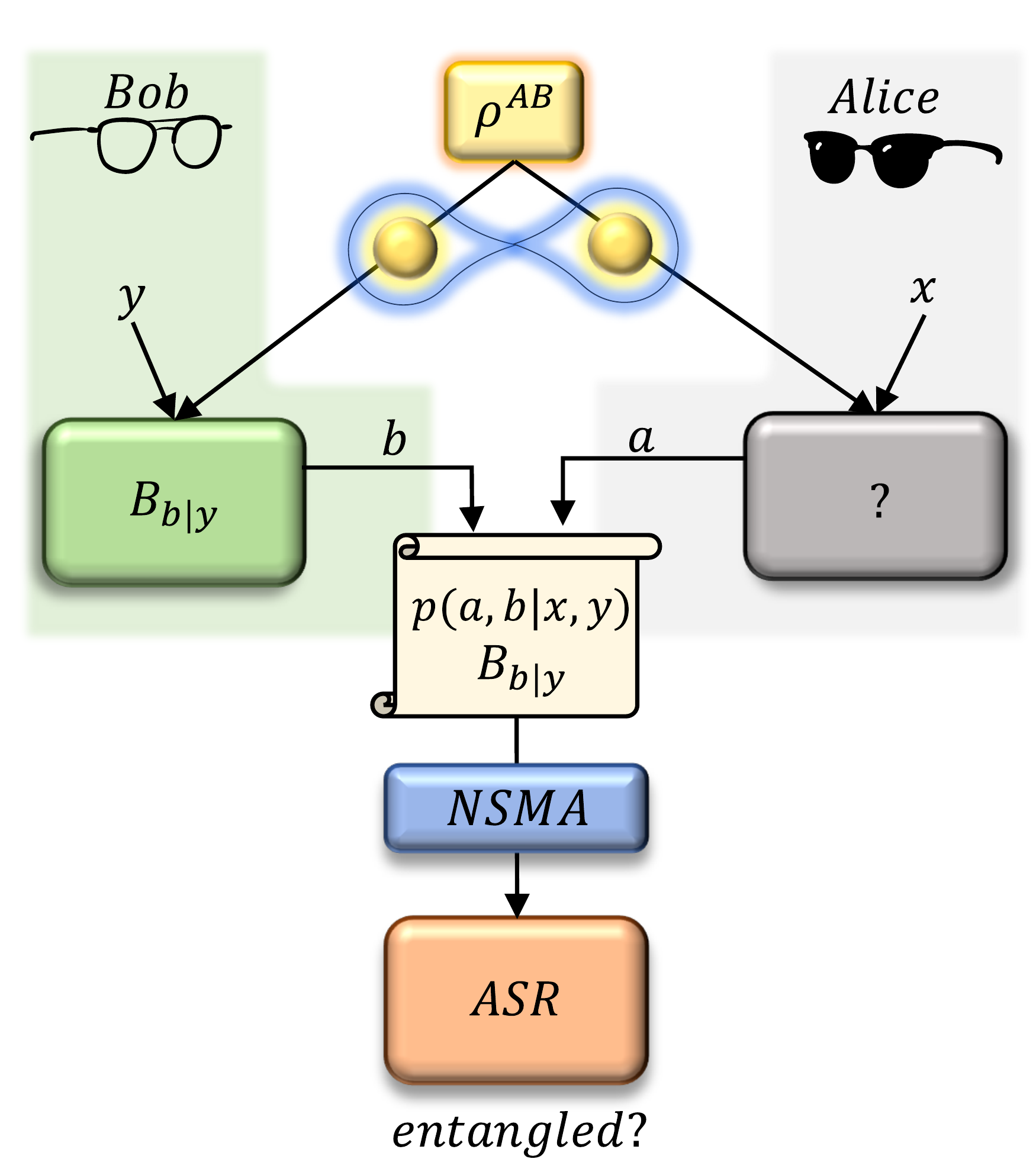}
     \caption{Adapted steering robustness protocol: A trusted party (Bob) wants to verify entanglement in a shared state $\rho^{AB}$ with an untrusted party (Alice) using measurement settings $y$ and $x$, respectively. Bob performs a POVM $B_{b|y}$, while Alice announces her measurement outcomes $a$. For statistical reasons, they repeat the experiment to form a joint probability distribution $\{\operatorname{p}(a,b|x,y)\}_{a,b,x,y}$. To correct for the effects of finite statistics, the NSMA is implemented, finding the closest non-signaling probability distribution that is realisable with Bob's measurements. The ASR is then calculated via a semidefinite program and used to verify entanglement.}
     \label{fig:prob_dist}
\end{figure}
Besides verifying if a shared state is steerable, one can determine how much noise, ubiquituous in practical demonstrations, can be added before a specific assemblage becomes unsteerable. This is quantified using the SR~\cite{Piani2015}, which is a measure of how much noise can be added to a state before it becomes unsteerable~\cite{Cavalcanti2016review}. Thus, any state with SR greater than zero is steerable.
To calculate SR, semidefinite programming (SDP)~\cite{skrzypczyk2014quantifying}, a convex optimization algorithm, over $\sigma_{a|x}$ is used to find the optimal witness,
\begin{align}
\begin{aligned}\label{EqSR}
\operatorname{SR} = \max _{\left\{F_{a \mid x}\right\}} &  \sum_{a, x} \operatorname{tr} \left[F_{a \mid x} \sigma_{a|x}\right]-1 \\
\text { s.t. } & \mathbb{1}-\sum_{a, x} D_{\lambda}(a \mid x) F_{a \mid x} \geq 0 \quad \forall \lambda \\
& F_{a \mid x} \geq 0 \quad \forall a, x ,
\end{aligned}
\end{align}
where $\{F_{a|x}\}_{a,x}$ is a set of Hermitian operators and $D_{\lambda}(\cdot|x)$ is a deterministic probability distribution. 

This method gives an optimal steering inequality, but it requires quantum state tomography on Bob's part of the state, demanding a precise calibration of measurement devices. Alternatively, we consider a less resource-intensive case, where Bob is limited to a fixed set of measurements, described by a POVM $\{B_{b|y}\}_{b}$ for each $y$, i.e. $\sum_b B_{b|y}=\openone$ and $B_{b|y}\geq0$ for each $b$, and collects his conditional probabilities to construct the joint probability distribution $\{\operatorname{p}(a,b|x,y)\}_{a,b,x,y}$. Therefore, 
we reformulate the steering robustness protocol using the joint probability distribution, instead of the state assemblage. A schematic figure of the protocol is presented in Fig.~\ref{fig:prob_dist}.

{\it Adapted steering robustness.---}
In our scenario, we have access to the joint probability distribution $\{\operatorname{p}(a,b|x,y)\}_{a,b,x,y}$, rather than $\sigma_{a|x}$, and define an adapted steering robustness (ASR):
\begin{align}
\begin{aligned} \label{ASR_good}
    \mathrm{ASR}&=\max_{\substack{ \{ \alpha_{x y}^{a b} \} }} \sum_{\substack{a, x \\
    b, y}} \alpha_{x y}^{a b} \mathrm{p}(a, b \mid x, y)-1 \\
    & \text { s.t. } \mathbb{1}-\sum_{a, x} D_\lambda(a \mid x) \sum_{b, y} \alpha_{x y}^{a b} B_{b \mid y} \geq 0 \quad \forall \lambda \\
    & \sum_{b, y} \alpha_{x y}^{a b} B_{b \mid y} \geq 0 \quad \forall a, x ,
\end{aligned}
\end{align}
where $\{\alpha_{x y}^{a b}\}_{a,b,x,y}$ are real coefficients and $B_{b|y}$ are Bob's trusted measurements. We note that there are no limits on the values $\{\alpha_{x y}^{a b}\}_{a,b,x,y}$ can take for a non-signaling scenario (Supplementary Material A \cite{supp}). 
The steering inequality, based on Eq.~\ref{ASR_good} is
\begin{equation}
\begin{aligned} \label{ASR_wit}
    \sum_{\substack{a, x \\ b, y}} \alpha_{x y}^{a b} \operatorname{p}(a,b|x,y) -1\leq 0,
\end{aligned}
\end{equation}
where $\{\operatorname{p}(a,b|x,y)\}_{a,b,x,y}$ are joint probabilities and $\{\alpha_{x y}^{a b}\}_{a,b,x,y}$ are real coefficients calculated via the SDP defined in Eq. \ref{ASR_good}.

Here, we consider a two-qubit state, i.e. a Werner state $\varrho_\mu = \mu \ket{\psi_s}\bra{\psi_s} + \frac{(1-\mu)}{4}\mathbbm{1}_4$, which is a probabilistic mixture of a maximally entangled singlet state $\ket{\psi_s} = \frac{1}{\sqrt{2}}(\ket{10}-\ket{01})$ with symmetric noise parameterised by the mixing probability $\mu \in [0,1]$~\cite{Werner1989}. We choose Alice and Bob's measurement settings to be $A_i = \vec{a}_i\cdot\vec{\sigma}$ and $B_i = \vec{b}_i\cdot\vec{\sigma}$, respectively, where $\vec{\sigma}=(\sigma_x,\sigma_y,\sigma_z)$ is a vector composed of the Pauli matrices, $\vec{a}_i\in\mathbb{R}^3$ and $\vec{b}_i\in\mathbb{R}^3$.  We chose Bob's measurement settings along $n$ random mutually unbiased bases (MUBs), quasi-uniformly (QU) distributed along the Bloch sphere (Supplementary Material B \cite{supp}). 

\begin{figure*}[t!]
    \centering
\includegraphics[width=\textwidth]{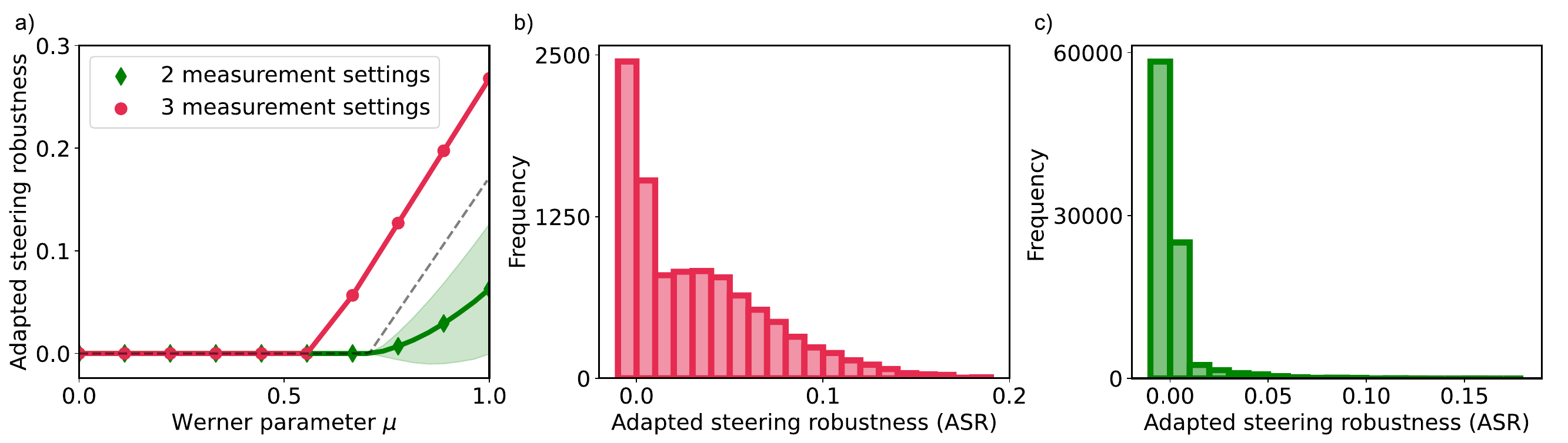}
\caption{(a) Simulated mean adapted steering robustness $\overline{ASR}$ as a function of mixing parameter $\mu$ for two (diamonds, green) and three (circles, red) random measurement settings per site. For three settings, Alice selects three orthogonal settings from a QU-random distribution, and Bob's measurements are fixed to ${\sigma_x, \sigma_y, \sigma_z}$. For two settings, Alice selects pairs of orthogonal settings from the same QU distribution, while Bob randomly selects pairs of Pauli measurements. Both cases are repeated for $n=50$ samples per data point. The data points represent the mean ASR values and one standard deviation (shaded area). For comparison, the dashed line shows the steering robustness, SR, for two measurement settings on Alice's side, chosen from the same QU distribution, and tomography using the Pauli measurements on Bob's side. Due to full state reconstruction by Bob, the SR achieves the maximum ASR value. (b) and (c) show histograms of adapted steering robustness (ASR) samples, computed using an experimental probability distribution after applying the NSMA algorithm. (b) Three settings case, 10,000 random mutually unbiased measurements, $75.5\%$ of the ASR values are non-zero, indicating steerability. (c) Two settings case, 90,000 pairs of random measurements, and $32.6\%$ of ASR are non-zero.
}\label{fig:ASR_hist_3_settings}
\end{figure*}

Firstly, we numerically simulate the ASR as a function of the mixing parameter $\mu$ (Fig.~\ref{fig:ASR_hist_3_settings}a). 
Our results demonstrate steering on average with positive $\overline{ASR}$ values for $\mu \geq \frac{1}{\sqrt{2}}$ and $\mu \geq \frac{1}{\sqrt{3}}$ for two and three measurement settings per side, respectively. These values match standard results for the SR when considering Werner states \cite{skrzypczyk2014quantifying}. 

For maximally entangled states with $\mu=1$ and three measurement settings per side, we obtain a maximum $ASR_{3}$ value of 0.2679, equivalent to the SR value. By reducing the measurements to two settings per side -- the minimal set size, we obtain $\overline{ASR}_{2} = 0.0627$ for $\mu=1$, when averaged over all combinations of measurements. For two perfectly aligned measurements per side, Eq.~(\ref{ASR_wit}) can take a maximum value of 0.1716. We note that in this scenario, the $ASR_{2}$ value is identical to Eq.~\ref{EqSR}. We also note that $SR_{2} \geq \overline{ASR}_{2}$ though $\max(ASR_{2})=SR_{2}$ as the SR has access to full tomographic information on Bob's side, so when measurements are random, SR will perform better, on average, than the ASR.

We implemented the steering protocols using a state-of-the art silicon photonic chip. The entangled photon pairs at 1550~nm are generated via Spontaneous Four Wave mixing in the two spiral-waveguide sources that are pumped by a 1550~nm fibre-couple continuous-wave laser with an output power (after fibre) of 300~mW. The generated state is verified using quantum state tomography at several stages throughout the experiment—in each case, we achieved a fidelity of ca. 93\% with the singlet state $\ket{\psi_{s}}$. Alice and Bob’s projective measurements are implemented by thermo-optic phase-shifters on chip together with coincidence detections using fibre-coupled superconducting nanowire single-photon detectors off-chip. See Fig. S1 in Supplementary Material B~\cite{supp} for a schematic of the chip.
The coincidence data was compiled into a joint probability distribution $\{\operatorname{p}_{\operatorname{exp}}(a,b|x,y)\}_{a,b,x,y}$ together with Bob's measurements $\{B_{b|y}\}_{b,y}$ to calculate the ASR (Eq.~\ref{ASR_good}). For this, we used an SDP encoded using CVX~\cite{cvx}, in addition to open-source codes from Ref.~\cite{cavalcanti2016Code} and Ref.~\cite{QETLABS}. The experimental data is available here~\cite{GitHub}.

The ASR, like any steering criterion, is ill-defined for signaling distributions. Applying the ASR algorithm to our experimental probability distribution leads to unbounded values for all random measurement settings in both the two and three measurements per side cases. It is noteworthy that many steering criteria can be evaluated even for signaling data without noticing such behaviour, although the fact that one uses signaling data would contradict the \textcolor{black}{constraints} that go into deriving the criterion in the first place. Here, the choice of the interval of the parameters $\{\alpha^{ab}_{xy}\}$ is such that it clearly highlights the problem that arises if one does not handle signaling data appropriately. Thus, we propose a method for finding a non-signaling approximation of the observed data, that is tailored for the steering setup.

{\it Correcting for signaling.---}
Quantum nonlocal correlations have to comply with the NS principle. Thus, the probability distributions of outputs of any subset of parties are independent of the remaining parties’ inputs.
Due to the finite nature of any experimental statistics, $\{\operatorname{p}_{\operatorname{exp}}(a,b|x,y)\}_{a,b,x,y}$ can never strictly satisfy NS conditions~\cite{renou2017inequivalence}. 
A probability distribution that exhibits signaling does not generate a valid assemblage and, hence, one should refrain from evaluating steering criteria on it. We emphasise that not satisfying the NS principle is different from closing well-known loopholes, such as the locality loophole \cite{wittmann2012loophole} and the detection loophole \cite{evans2014optimal}. However, we highlight that even in a loophole-free experiment, the effects of finite statistics remain and must be corrected.

Previous methods of finding the closest NS probability distribution minimised the relative entropy between a given signaling distribution and an NS distribution~\cite{cavalcanti2016quantitative}. The NS distribution found is physical but it may not be realisable with Bob's measurements $B_{b|y}$. Thus, we propose a \textcolor{black}{one-sided} device independent non-signaling model algorithm (NSMA) that finds the closest NS probability distribution by minimising the amount of signaling $t \geq 0$ one needs to add to a signaling distribution, given $B_{b|y}$, to generate an NS distribution. The algorithm is defined by an SDP:
\begin{align} \label{closest_NS_SDP}
\begin{aligned}
t =\min_{\left\{\tilde{\sigma}_{a|x}\right\}} \quad &  \frac{1}{\#x}\sum_{a,x} \operatorname{tr}\left[ \tilde{\sigma}_{a|x} \right] -1 \\
\text { s.t. } & \operatorname{tr}\left[ \tilde{\sigma}_{a|x}B_{b|y} \right] - \operatorname{p}_{\operatorname{exp}}(a,b|x,y)  \geq 0 \quad \forall a,b,x,y \\
& \tilde{\sigma}_{a|x}  \geq 0 \quad \forall a, x \\
& \sum_{a} \tilde{\sigma}_{a|x} = \sum_{a} \tilde{\sigma}_{a|x'} \quad \forall x\neq x',
\end{aligned}
\end{align}
where $\#x$ is the number of Alice's measurement settings and $\tilde{\sigma}_{a|x}=(1+t)\sigma_{a|x}$, where $\sigma_{a|x}$ is the NS assemblage we are searching for. 
The corresponding NS probability distribution is $\operatorname{p}_{\operatorname{exp}}^{\operatorname{NS}}(a,b|x,y) = \frac{\operatorname{tr}\left[ \tilde{\sigma}_{a|x} B_{b|y} \right]}{1+t},$
see Supplementary Material C \cite{supp} for further details on the derivation of the NSMA. We then ensure the NSMA is not biased towards indicating steerability - in that its effects do not increase the probability of an $ASR>0$. We used Monte Carlo methods to simulate the impact of finite statistics followed by the NSMA on the ASR for sets of states and measurements on the boundary conditions of the LHS model. We found that the NSMA underestimates the steerability here. See Supplementary Material D \cite{supp} for details. 

We applied this method to our experimental probability distribution $\{\operatorname{p}_{\operatorname{exp}}(a,b|x,y)\}_{a,b,x,y}$ to find the closest NS distribution $\{\operatorname{p}_{\operatorname{exp}}^{\operatorname{NS}}(a,b|x,y)\}_{a,b,x,y}$, for each of the 10,000 randomised measurements. Fig.~\ref{fig:ASR_hist_3_settings}b shows the distribution of ASR values for three measurement settings per side. Our experimental results show a $75.5\%$ chance of violating the ASR inequality. We find $\overline{ASR}_{exp-3} =0.033$ and $0 \leq ASR_{exp-3} \leq 0.1885$. For two measurement settings per side, we consider every combination of pairs of the three MUB measurements. This results in 90,000 samples and $\overline{ASR}_{exp-2} = 0.0049$, with values $0 \leq ASR_{exp-2} \leq 0.1632$ for all samples (Fig.~\ref{fig:ASR_hist_3_settings}c). We violate the ASR-2 inequality with a probability of $32.6\%$.

{\it Comparison to other steering inequalities.---}
To assess the strength of the ASR, we calculated the probability of violation of the NS distribution for other well-known steering inequalities: linear-steering (LS) \cite{Cavalcanti2009,Saunders2010}, CHSH-like steering (CHSH-LS) \cite{Cavalcanti2015}, dimension-bounded steering (DBS) \cite{Moroder2016} and rotationally-invariant steering (RIS) \cite{Wollmann2018}  (Supplementary Material E \cite{supp}). In Fig.~\ref{bar_plot}, we compare the ASR inequality violation probability (Eq.~\ref{ASR_wit}) to comparable steering inequalities. In Supplementary Material F \cite{supp}, we present a table comparing the probabilities of violation before and after applying the NSMA to highlight the importance of correcting for the effects of finite statistics.

We find the probability of violating the ASR inequality greatly outperforms all other considered inequalities. In the three settings case, for the rotation-invariant inequalities, DBS-3 and RIS-3, the NS experiment only achieves $6.4\%$ and $11.3\%$ probability of violation respectively and the linear inequality LS-3 performed the worst, having a $0.3\%$ probability of violation. For two measurements, the probability of demonstrating steering using the other inequalities ranges between $1.1\%$ for LS-2 and $10.7\%$ for CHSH-like steering. Numerical simulations comparing the probability of violation for a maximally entangled state ($\mu=1$) can be found in Supplementary Material F \cite{supp}.

\begin{figure}
\centering
 \begin{tikzpicture}
 \begin{axis}[
ybar=3pt,
bar width=10.5pt,
height=6.45cm,
width=0.49\textwidth,
xmin=1.2,
xmax=4.5,
ymin=6,
ymax=100,
enlarge x limits = {abs= 0.75},
enlarge y limits = {abs = 9.0},
xtick=data,
ylabel=Probability of violation ($\%$),
y label style={at={(axis cs:-0.1,0)},anchor= 
west},
y label style={scale=1.12},
x tick label style={rotate=0},
x label style={scale=1.12},
y tick label style={/pgf/number format/1000 sep=},
xticklabels={ASR, LS, DBS, RIS, CHSH-LS},
legend entries= {2 measurement settings, 3 measurement settings},
legend style={at={(1,1)},anchor=north east}
]

\addplot  [draw = {green},
    line width = 0.75mm,
    draw opacity=0.95,
    fill = green,
    fill opacity=0.4,
    text opacity=1
]coordinates { (1,32.56) (2,1.12)  (3,2.30) (4,4.99) (4.9,10.73) };

\addplot [draw = {awesome},
    line width = .75mm,
    draw opacity=0.95,
    fill = awesome,
    fill opacity=0.34,
    text opacity=1
] coordinates { (1,75.5)  (2,0.3)  (3,6.4) (4,11.3) };

\node[above][scale=1.1] at (axis cs:0.75, 32.8) {\footnotesize 32.6};
\node[above][scale=1.1] at (axis cs:1.15, 75.7) {\footnotesize 75.5};

\node[above][scale=1.1] at (axis cs:1.8, 1.3) {\footnotesize 1.1};
\node[above][scale=1.1] at (axis cs:2.15, 0.5) {\footnotesize 0.3};

\node[above][scale=1.1] at (axis cs:2.8, 2.5) {\footnotesize 2.3};
\node[above][scale=1.1] at (axis cs:3.15, 6.6) {\footnotesize 6.4};

\node[above][scale=1.1] at (axis cs:3.8, 5.2) {\footnotesize 5.0};
\node[above][scale=1.1] at (axis cs:4.168, 11.5) {\footnotesize 11.3};

\node[above][scale=1.1] at (axis cs:4.7, 10.9) {\footnotesize 10.7};

 \end{axis}
\end{tikzpicture}
\caption{Experimental probabilities of violation for two (green) and three (red) orthogonal measurements per site for different criteria. Alice's measurement settings are chosen from a quasi-uniform random distribution on the surface of the Bloch sphere, whilst Bob's measurement settings are fixed along ${\sigma_x, \sigma_y,\sigma_z}$. The probabilities of violation are calculated using the non-signaling (NS) probability distribution, closest to our experimental probabilities, which is obtained using the NSMA. We compare the ASR probabilities of violation to comparable inequalities: linear steering (LS), dimension-bounded steering (DBS), rotationally-invarient steering (RIS) and Clauser-Horne-Shimony-Holt-like steering (CHSH-LS).
}
\label{bar_plot}
\end{figure}

{\it Discussion.---}
Our approach to quantifying steering represents an important step towards certifying entanglement for future quantum networks. By harnessing the power of SDPs, we have demonstrated the noise-robustness of our ASR witness. Furthermore, we have shown our approach performs particularly well in scenarios involving fully misaligned measurements and finite statistics, where other inequalities under perform. 

In previous experiments, the effects of finite statistics have often been excluded or additional assumptions have been made~\cite{giustina2015,shalm2015,hensen2015, rosenfeld2017}. Without satisfying the non-signaling \textcolor{black}{constraint}, any test that excludes local hidden models cannot genuinely prove that the observed entanglement is not a result of signaling. Here, in our proof-of-principle demonstration, we witness entanglement in a steering scenario, with statistics that are by definition of the NSMA algorithm – non-signaling. Therefore, we are not required to exclude any data or make any additional assumptions as is often the case.

Additionally, our protocols require minimal assumptions about either party's measurements. On one hand, removing the constraint of full tomography on the trusted party allows us to present optimised inequalities based solely on correlations. On the other hand, we retain the loss and noise robustness of steering protocols. Furthermore, our NSMA is an SDP based method for fitting data into a quantum model. Albeit the method here is presented for quantum steering, it is directly adaptable to quantum state estimation and quantum algorithms~\cite{philip2023}. We expect that such a convex-geometric linear method can be favourable over distance-based and entropy-based methods in terms of computational efficiency and will contribute to the emerging field of finite statistics in quantum networks. Overall, the potential for using experimental techniques to explore the effects of finite statistics is an exciting area of research with many possible applications in quantum information processing~\cite{lee2018} and other fields~\cite{barrett2005,acin2006}.

{\it Acknowledgments.---}
We thank Tiff Brydges, Benoît Vermersch, Paul Skrzypczyk, Josh Silverstone, and Jeremy Adcock for useful discussions. We thank Graham Marshall, Callum Wilkes, and Laurent Kling for their contributions to the design and characterisation of the silicon photonic chip.
This project was supported by the Centre for Nanoscience and Quantum Information (NSQI) and by funding from the European Union’s Horizon 2020 research and innovation programme under the Marie Skłodowska-Curie grant agreement No 892242, VERIqTAS-QuantERA II Programme under Grant Agreement No 101017733, the Polish National Science Center (grant No 2021/03/Y/ST2/00175),
and EPSRC grants (EP/T00097X/1) QUANTIC, EP/M024385/1, and EP/SO23607/1, and EP/W003252/1, as well as from CAPES/Brazil and the Swiss National Science Foundation (Ambizione PZ00P2- 202179). J.W. acknowledge support from the Natural Science Foundation of China (nos. 62235001, 61975001), the Innovation Program for Quantum Science and Technology (2021ZD0301500), the National Key R\&D Program of China (2019YFA0308702), Beijing Natural Science Foundation (Z190005, Z220008). X.Q. acknowledges support from the National Natural Science Foundation of China (62075243). J.L.O. acknowledges funding from Royal Academy of Engineering Chair in Emerging Technologies. A.C.S.C. was supported by CNPq/Brazil, Grants No. 308730/2023-2 and No. 409673/2022-6. A.C.O.J. was supported by NCN SONATA
BIS Project No. 2022/46/E/ST2/00115.

Correspondence and requests for materials should be addressed to the corresponding authors.

\bibliographystyle{apsrev4-2}
\bibliography{refs.bib}

\widetext
\setcounter{equation}{0}
\setcounter{figure}{0}
\setcounter{table}{0}
\setcounter{page}{1}
\setcounter{section}{0}
\makeatletter
\renewcommand{\theequation}{S\arabic{equation}}
\renewcommand{\thefigure}{S\arabic{figure}}
\renewcommand{\bibnumfmt}[1]{[#1]}
\renewcommand{\citenumfont}[1]{#1}

\section{Supplemental Material: Correcting for finite statistics effects in a quantum steering experiment}

\subsection{A. Deriving the adapted steering robustness} \label{derivation_ASR}

In this appendix, we show the derivation of the ASR (Eq.2 in the main text). Firstly, we note that we are interested in the scenario in which our untrusted party (Alice) performs measurements, denoted $x$, with measurement outcomes $a$. Our trusted party (Bob) performs measurements, denoted $y$, with measurement outcomes $b$. As we trust Bob, we also know the positive measurement operators $\{B_{b|y}\}_{b,y}$. Note that, unlike in the standard steering scenario, there is no requirement for Bob's measurements to provide full tomographic information. Alice and Bob perform a set of measurements, resulting in the set of joint probabilities $\{\operatorname{p}(a,b|x,y)\}_{a,b,x,y}$. We can define this set of joint probabilities in terms of the familiar state assemblage, $\sigma_{a|x}$ and Bob's measurement operators $B_{b|y}$ as such:
\begin{align}
\begin{aligned}
    \operatorname{p}(a,b|x,y) &= \operatorname{tr}\left[\left(A_{a|x} \otimes B_{b|y}\right) \rho_{AB}\right] \\
    & = \operatorname{tr}\left[ \operatorname{tr}_{\mathrm{A}}\left[\left(A_{a \mid x} \otimes \mathbb{1}\right) \rho_{AB}\right] B_{b|y}\right] \\
    & = \operatorname{tr}\left[ \sigma_{a|x} B_{b|y}\right].
\end{aligned}
\end{align}
In order to generate a function to be optimised in a semidefinite program (SDP), we sum the probabilities, weighting them with real coefficients $\alpha_{xy}^{ab}$:
\begin{align} \label{SR_Adapted}
\begin{aligned}
    \sum_{\substack{a, x \\ b, y}} \alpha_{x y}^{a b} \operatorname{p}(a,b|x,y) &=
    \sum_{\substack{a, x \\ b, y}} \alpha_{x y}^{a b} \operatorname{tr} \left[\sigma_{a|x} B_{b|y} \right]  \\
    &= \sum_{a,x} \operatorname{tr}\left[ \sigma_{a|x} \sum_{b,y} \alpha_{x y}^{a b} B_{b|y}   \right] \\
    & = \sum_{a,x} \operatorname{tr}\left[ \sigma_{a|x} \tilde{F}_{a|x}   \right],
\end{aligned}
\end{align}
where we define a new Hermitian operator $\tilde{F}_{a|x} := \sum_{b,y} \alpha_{x y}^{a b} B_{b|y} $. 

We use Eq.~\ref{SR_Adapted} to define a semidefinite program inspired by the steering robustness that fits the quantum model pertaining to Bob's measurements and NS:
\begin{align}
\begin{aligned} \label{ASR}
\operatorname{ASR}\left[\sigma_{a|x}, B_{b|y}\right]=\max _{\left\{\alpha_{x y}^{a b}\right\}} & \sum_{a,x} \operatorname{tr}\left[ \tilde{F}_{a|x} \sigma_{a|x}    \right] -1 \\
\text { s.t. } & \mathbb{1}-\sum_{a x} D_{\lambda}(a \mid x) \tilde{F}_{a \mid x} \geq 0 \quad \forall \lambda \\
& \tilde{F}_{a \mid x} \geq 0 \quad \forall a, x .
\end{aligned}
\end{align}
 Here, the only difference between ASR and the standard formulation of SR is the definition of $\tilde{F}_{a|x}$. We are restricting $\tilde{F}_{a|x}$ to be equal to Bob's settings $B_{b|y}$ multiplied by some real constant $\alpha_{x y}^{a b}$. This means the semidefinite program no longer optimises over the set of Hermitian matrices $\{F_{a|x}\}_{a,x}$, but rather over the set of real coefficients $\{\alpha_{x y}^{a b}\}_{a,b,x,y}$. 

Finally, we can rewrite the ASR in the form it appears in the main text by substituting in the definition of $\tilde{F}_{a|x}$:
\begin{align} 
\begin{aligned}
\operatorname{ASR}\left[\operatorname{p}(a,b|x,y), B_{b|y}\right]=\max _{\{\alpha_{x y}^{a b}\}}  &\sum_{\substack{a, x \\ b, y}} \alpha_{x y}^{a b} \operatorname{p}(a,b|x,y) -1 \\
\text { s.t. } & \mathbb{1}-\sum_{a x} D_{\lambda}(a \mid x) \sum_{b,y} \alpha_{x y}^{a b} B_{b|y}  \geq 0 \quad \forall \lambda \\
& \sum_{b,y} \alpha_{x y}^{a b} B_{b|y}  \geq 0 \quad \forall a, x
\end{aligned}
\end{align}
We note that we place no conditions on the values of the real coefficients, which constitute the variables maximised over in the SDP. As long as the constraint that the Hermitian operators $\tilde{F}_{a|x}$ are positive, we still satisfy the necessary conditions of the SDP. Importantly, we must also ensure the joint probability distribution $\{\operatorname{p}(a,b|x,y)\}_{a,b,x,y}$ satisfies three principles: non-negativity, non-signaling and normalisation. See supplementary material~\ref{SDP_supp_mat} for details on our method to satisfy the non-signaling requirement.

\subsection{B. Experimental set-up}\label{experiment}
\begin{figure}[htbp]
\includegraphics[width=0.8\textwidth]{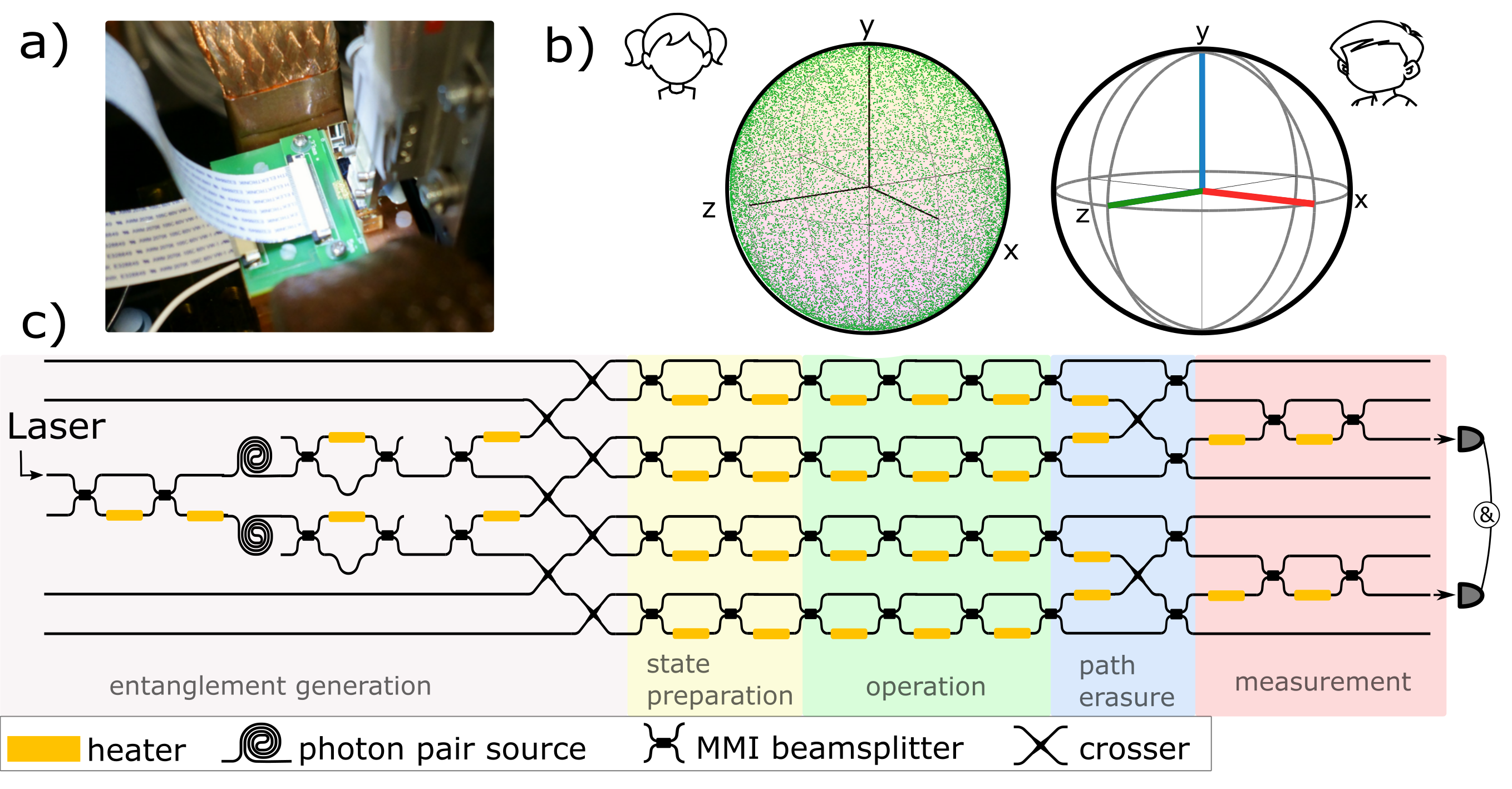}
\caption{A photograph of the silicon photonic chip used in the experiment (a). The distribution of the measurement settings over the Bloch sphere. Whilst Bob's settings are fixed along $\sigma_{x}$, $\sigma_{y}$, and $\sigma_{z}$, Alice can perform random orthogonal measurements chosen from a quasi-random distribution (b). A schematic of the experimental layout is depicted in (c). The device includes five functional parts that collectively implement a given $SU(4)$ operation. Part I (purple) is the generation of a path-entangled ququart state, Part II is preparing the initial single-qubit states as the input into the gate (yellow), Part III implements the linear terms $A_i$ and $B_i$ (Eq. \ref{eq:chipU}) of the $SU(4)$ operation (green), Part IV (blue) realizes the linear combination of terms $A_i \otimes B_i$ together with post-selection, and Part V (red) part rotates the output state so that it can be measured at the desired basis before it is routed off chip for detection.}
\label{fig0}
\end{figure}

The silicon photonic chip used for the experiment is described in Fig.\ref{fig0}~\cite{Qiang2018}. It consists of 4 spiral-waveguide spontaneous four-wave mixing (SFWM) photon-pair sources~\cite{silverstone2014}, 4 laser pump rejection filters, 82 multi-mode interferometer (MMI) beamsplitters, and 56 simultaneously operating thermo-optic phase shifters~\cite{silverstone2014}. The chip is operated using an external electrical control, laser input, and fiber-coupled superconducting single-photon nanowire detectors.

To create photon pairs, the four SFWM sources are pumped with a laser, which is split across the four sources according to complex coefficients $\alpha_i$. The resulting spatially bunched photon pairs are generated coherently in any one of the four sources. By post-selecting when signal and idler photons exit at the top two output modes (qubit 1) and the bottom two (qubit 2), respectively, a path-entangled ququart state, $\ket{\Phi}=\alpha_0 \ket{1}_a\ket{1}_e + \alpha_1 \ket{1}_b\ket{1}_f + \alpha_2 \ket{1}_c\ket{1}_g + \alpha_3 \ket{1}_d\ket{1}_h$, can be obtained via post-selection at the end of the `entanglement generation' stage.
To generate an arbitrary two-qubit unitary $U\in \mbox{SU}(4)$ on this photonic chip, the appropriate decomposition into a linear combination of four terms: $U = \sum_{i=0}^3 {\alpha_i A_i \otimes B_i}$ has to be found. Here $A_i$ and $B_i$ are single-qubit gates and $\alpha_i$ are complex coefficients satisfying $\sum_{i=0}^3 \left|\alpha_i \right|^2=1$, and thus implements universal two-qubit processing via two-photon ququart entanglement~\cite{Qiang2018}.
To realise this arbitrary unitary on the chip, each spatial mode after `entanglement generation' is extended into two levels to form qubits $\ket{\varphi_1}$ or $\ket{\varphi_2}$, and $A_i$ and $B_i$ are applied to $\ket{\varphi_1}$ and $\ket{\varphi_2}$ respectively, evolving $\ket{\Phi}$ into $\sum_{i=0}^3 {\alpha_i}A_i \ket{\varphi_1}_{u^i} B_i\ket{\varphi_2}_{v^i}$, where $u^i \in\{a,b,c,d \}$ and $v^i \in \{e,f,g,h \}$. By combining the qubits $a,b,c,d$ into one final-stage qubit, and the qubits $e,f,g,h$ into the second, it yields the state as 
\begin{align}
\sum_{i=0}^3 {\alpha_i A_i \otimes B_i}\ket{\varphi_1}\ket{\varphi_2}
\label{eq:chipU}
\end{align}
after path information erasure, realizing the target unitary $U$. Overall, the chip is designed for the generation of high-dimensional entangled photon states, which have potential applications in quantum communication and quantum computing.

For the photon pair generation in the experiment, we pumped the photon sources using a continuous-wave laser at 1550nm with an output power of 300 mW (after amplification). Entanglement is generated by implementing a high-fidelity CNOT gate by decomposing it to linear single-qubit gate combinations to realize the maximally entangled-target state $|\Psi^{-1}\rangle=\frac{1}{\sqrt{2}}(\ket{01}- \ket{10})$. In fact, we require only the linear combination of two terms $U=\alpha_0 A_0 \otimes B_0+ \alpha_1 A_1 \otimes B_1$ to implement  a CNOT gate on the chip, which thus simplifies the required functional area of the full device as shown in Fig.~\ref{fig0}(c).

We repeatedly performed a tomographic reconstruction to ensure the state quality is sufficiently high. We found our state quality to have $\sim 92\%$ fidelity with a maximally entangled state. We attribute this to environmental fluctuations.
Finally, the qubits are spatially recombined to implement Alice's and Bob's
$m$ measurement directions with projective measurements before detection. Bob's measurements are limited to the Pauli measurements ${\sigma_x, \sigma_y, \sigma_z}$, whilst Alice performs random orthogonal measurements. For experimental reasons, we chose a quasi-uniform (QU) random distribution instead of a truly uniform random distribution on the Bloch sphere (Fig.~\ref{fig0}a). 
This allowed us to significantly reduce the number of performed measurements in total to 240600, and thus our data collection time to 167 days.

\subsection{C. Derivation of non-signaling algorithm} \label{SDP_supp_mat}
For practical reasons, experiments rely on a finite number of samples to estimate a probability distribution. Even with a large sample size, we are still restricted by the effects of finite statistics. As demonstrated in our experiment, this results in an experimental probability distribution that does not completely satisfy the non-signaling principle. Often this effect is not taken into account or additional assumptions have to be made resulting in collected data being excluded. Importantly, the semidefinite program used to calculate the ASR highlights the effects of finite statistics, as the maximisation problem becomes unbounded, whilst still satisfying the other constraints in the program. Thus, demanding either a perfect, non-signaling probability distribution, or a correction of this effect. 

In this appendix, we outline the non-signaling model algorithm (NSMA) that can numerically identify a signaling-free probability distribution as defined in the main text. The idea is as such: given some signaling probability distribution, we want to find the closest NS probability distribution that can be generated from a valid steering assemblage. Previous work achieved this by minimising the relative entropy~\cite{cavalcanti2016quantitative}. However, we trust Bob's measurement settings $\{B_{b|y}\}_{b,y}$. Therefore, we impose an additional constraint that the NS probability distribution is physically realisable with Bob's measurement settings.

\begin{figure}[htbp]
    \centering
    \includegraphics[width=0.5\textwidth]{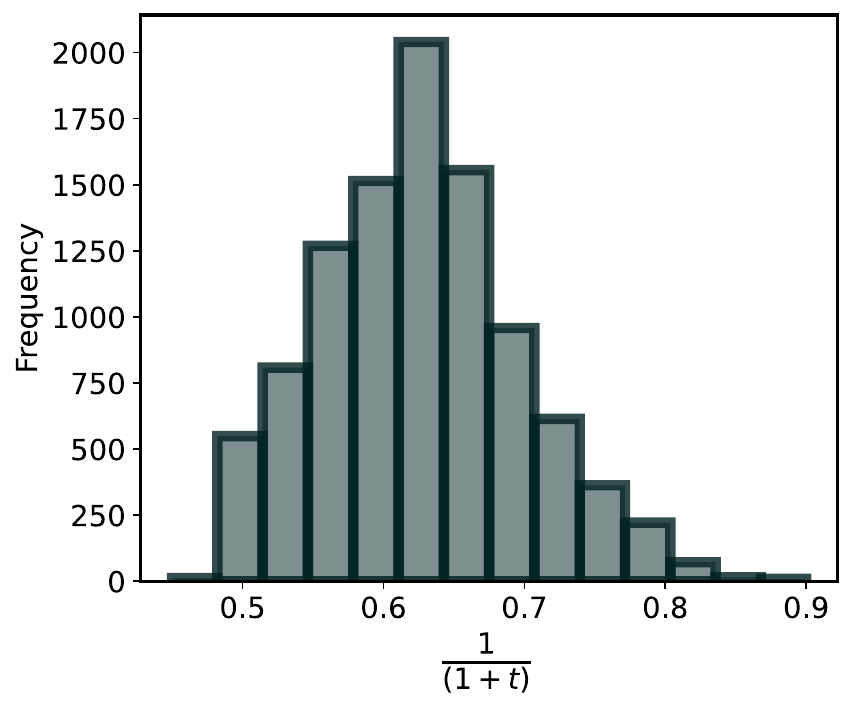}
    \caption{Histogram of 10,000 signaling robustness values $t$, between the three settings per side experimental probability distribution and the closest NS probability distribution, realisable with Bob's measurement settings. }
    \label{fig:distances}
\end{figure}

In such a scenario, an SDP inspired by the steering robustness can be used. We define the set of all physically realisable probability distributions with $\{B_{b|y}\}_{b,y}$ as $\mathcal{R}$.
We define the SDP in its primal form as:
\begin{align} 
\begin{aligned}
\min t \geq 0 \\
\text { s.t. } & \frac{\operatorname{p}_{\operatorname{exp}}(a,b|x,y)+t\tilde{\operatorname{p}}(a,b|x,y)}{1+t} = \operatorname{tr}\left[ 
 \sigma_{a|x}B_{b|y} \right] \in \mathcal{R} \\
& \tilde{\operatorname{p}}(a,b|x,y)  \geq 0 \quad \forall a,b,x,y \\
& \sum_{a,b} \tilde{\operatorname{p}}(a,b|x,y) = 1 \quad \forall x,y,
\end{aligned}
\end{align}
where $t$ is the signaling robustness that we wish to minimise. Rearranging for $\tilde{\operatorname{p}}(a,b|x,y)$ and defining $\tilde{\sigma}_{a|x}=(t+1)\sigma_{a|x}$, we arrive at the dual form of the SDP (as seen in the main text):
\begin{align} 
\begin{aligned}
t =\min_{\left\{\tilde{\sigma}_{a|x}\right\}} \quad &  \frac{1}{\#x}\sum_{a,x} \operatorname{tr}\left[ \tilde{\sigma}_{a|x} \right] -1 \\
\text { s.t. } & \operatorname{tr}\left[ \tilde{\sigma}_{a|x}B_{b|y} \right] - \operatorname{p}_{\operatorname{exp}}(a,b|x,y)  \geq 0 \quad \forall a,b,x,y \\
& \tilde{\sigma}_{a|x}  \geq 0 \quad \forall a, x \\
& \sum_{a} \tilde{\sigma}_{a|x} = \sum_{a} \tilde{\sigma}_{a|x'} \quad \forall x\neq x',
\end{aligned}
\end{align}
We apply this SDP to each $\operatorname{p}_{\operatorname{exp}}(a,b|x,y)$ in our set of 10,000 experimental distributions. The NS probability distributions, $\operatorname{p}_{\operatorname{exp}}^{\operatorname{NS}}(a,b|x,y) = \frac{\operatorname{tr}\left[ \tilde{\sigma}_{a|x} B_{b|y} \right]}{1+t}$, are then extracted from the SDP. We assume that the NS distribution found represents the underlying experimental distribution. The distribution of signaling robustness values $t$, can be seen in Fig.~\ref{fig:distances}. 

\subsection{D. Investigation of bias of the NSMA} \label{NSMA_biasedness}

In this appendix, we analyse the NSMA to investigate any potential bias it may generate. To begin, we consider an ideal experimental scenario in which a fixed shared state is perfectly  measured repeatedly, to build up statistics. Due to the finite number of rounds in any experiment, it is not possible to exactly reproduce the true underlying non-signaling (NS) probability distribution $p(a,b|x,y)$. Instead, Poissonian noise in the sampling generally results in apparent signaling which we then correct with the NSMA algorithm.

Similarly to the steering robustness, the NSMA algorithm finds the smallest amount of signaling distribution that needs to be added to the apparent signaling distribution to make it NS. Once this has been found, the NS distribution is constructed and we assume this represents the true underlying statistics. However, the manifold of NS distributions is not convex and so a valid question to ask is how close the probability distribution identified by the NSMA is to the true underlying distribution. Further, we are interested to know if our algorithm is biased in the sense that it identifies probability distributions that are more steerable. This could possibly lead to the NSMA finding a steerable NS distribution when the true underlying distribution was actually unsteerable. Further, such instances of false positives would lead to higher probabilities of violations than the true distribution would give, biasing us in the steerable direction.

Here, we address this question by simulating the effects of sampling in finite rounds from a distribution using Monte Carlo methods.
For this, we choose states and measurements that are on the surface of unsteerable distributions which can be described by local hidden state models. These distributions have nearby steerable distributions and thus, are more likely to lead to projections onto steerable distributions. We simulate the ideal probability distributions and calculate the ASR for the following three scenarios:
\begin{enumerate}
    \item A maximally entangled state, where Bob and Alice perform two orthogonal measurements each but in orthogonal planes, e.g. Alice measures $\sigma_{x}$, $\sigma_{y}$ and Bob measures $\sigma_{y}, \sigma_{z}$.
    \item A Werner state with $\mu=\frac{1}{\sqrt{2}}$ and two perfectly correlated measurements, e.g. Alice and Bob both measure $\sigma_{x}$ and $\sigma_{y}$.
    \item A Werner state with $\mu=\frac{1}{\sqrt{3}}$ and three perfectly correlated measurements, e.g. Alice and Bob both measure $\sigma_{x}$, $\sigma_{y}$ and $\sigma_{z}$.
\end{enumerate}
 For all these scenarios, we have $ASR=0$, thus they are unsteerable. Then, using Poissonian statistics, we simulate $N$ experiments each sampling a finite number of times, $n$, from the underlying distribution. This results in $N$ probability distributions that all contain some amount of apparent signaling due to finite statistics. For each of the $N$ simulated experiments, we then apply the NSMA, followed by the ASR. As we know, the true underlying state has $ASR=0$, therefore any of the $N$ experiments that return an $ASR>0$ are examples of false positives. The distribution of $ASR$s, in particular the number that return false positives, $ASR>0$, allows us to infer the effect that finite statistics and the NSMA have on our ability to detect steering and our confidence in these results. We note that we consider here only the effects of finite statistics, and exclude the effects of systematic noise from measurement imperfections.

To apply our simulations to realistic conditions, we consider a number of counts, $n$, that consist of a representation of coincidence count rates for the chosen experimental settings, as well as accidental counts generated by noise. Based on our silicon photonics platform, we take the average number of coincidence counts (samples) as $n=500$. The larger we choose $n$ in the simulation, the better we approximate the true distribution and so the smaller the amount of apparent signaling. Therefore, a small sample size will lead to low confidence in any witnessed steerability. For each of the $N$ simulated experiments, we assume the quantum states and their measurements have a true underlying statistics following a Poissonian distribution with corresponding mean.

\begin{figure*}[h!]
    \centering
\includegraphics[width=0.98\textwidth]{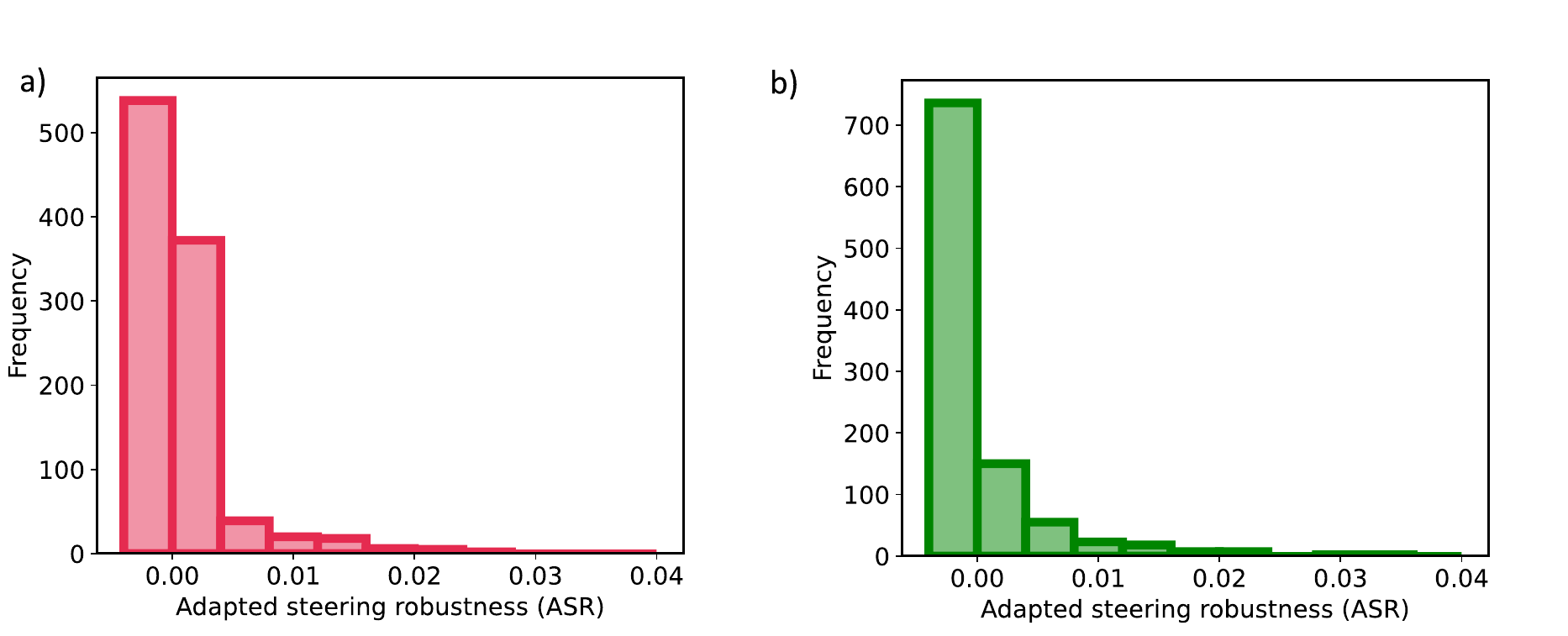}
\caption{
Histogram of adapted steering robustness (ASR) values for 1000 simulated perfect experiments, computed using Monte Carlo methods. The average number of coincidence counts is $n=500$. (a) Scenario 3: a Werner state with $\mu=\frac{1}{\sqrt{3}}$ and three perfectly correlated measurements.  (b) Scenario 2: a Werner state with $\mu=\frac{1}{\sqrt{2}}$ and two perfectly correlated measurements. 
}\label{fig:NSMA_biasedness_hists}
\end{figure*}

We simulate $N=1000$ experiments, with $n=500$ counts for each of the three scenarios defined earlier. If we expected the NSMA to cause the steering witness values to be normally distributed around the true value ($ASR=0$), then we would expect $\approx 50\%$ of the 1000 experiments to result in an $ASR>0$. Instead, we find that scenario 1 has $15.4\%$ $ASR>0$, scenario 2 has $26.4\%$ $ASR>0$ and scenario 3 has $46.2\%$ $ASR>0$, indicating that NSMA and ASR tend to cause a decrease in the observed steering. Histograms of the ASR values for scenarios 2. and 3. can be seen in Fig. \ref{fig:NSMA_biasedness_hists}.

We then calculate the ASR value $x$ within a 95$\%$ confidence window of our Monte Carlo simulation. Thus, there is a $5\%$ chance that a false positive will have an $ASR>x$. 
We found that for scenario 1, $95\%$ of our Monte Carlo simulated experiments had an $ASR < 4.5 \times 10^{-5}$. For scenario 2, the value was $x=0.0088$ and for scenario 3, the value was $x=0.0086$. We apply these $x$-values to scenarios 2 and 3 and apply them as false-positive bounds to our experimental results. For the three settings case, we find that $61.34\%$ of the ASR values are greater than 0.0086. For the two settings case, we find that $12.55\%$ of the ASRs values are greater than 0.0088. In order to make a fair comparison with the other steering inequalities, we apply the same 95$\%$ confidence window to them. A histogram comparing the probability of violation with a 95$\%$ confidence window can be seen in Fig. \ref{bar_plot_95_confidence_window}. Our results show the advantages of our ASR protocol in comparison to other protocols. 

\begin{figure}
\centering
 \begin{tikzpicture}
 \begin{axis}[
ybar=3pt,
bar width=10.5pt,
height=6.45cm,
width=0.8\textwidth,
xmin=1.2,
xmax=4.5,
ymin=6,
ymax=65,
enlarge x limits = {abs= 0.75},
enlarge y limits = {abs = 9.0},
xtick=data,
ylabel=Probability of violation ($\%$),
y label style={at={(axis cs:-0.1,0)},anchor= 
west},
y label style={scale=1.12},
x tick label style={rotate=0},
x label style={scale=1.12},
y tick label style={/pgf/number format/1000 sep=},
xticklabels={ASR, DBS, RIS, LS, CHSH-LS},
legend entries= {2 measurement settings, 3 measurement settings},
legend style={at={(1,1)},anchor=north east}
]

\addplot  [draw = {green},
    line width = .75mm,
    draw opacity=0.95,
    fill = green,
    fill opacity=0.4,
    text opacity=1
]coordinates { (1,12.55)   (2,1.4411) (3,4.6789) (4, 1.07) (4.9,10.10)  };

\addplot [draw = {awesome},
    line width = .75mm,
    draw opacity=0.95,
    fill = awesome,
    fill opacity=0.34,
    text opacity=1
] coordinates { (1,61.34)  (2,0.44)  (3,10.93) (4,0.23) };

\node[above][scale=1.1] at (axis cs:0.85, 13) {\footnotesize 12.6};
\node[above][scale=1.1] at (axis cs:1.075, 62) {\footnotesize 61.3};


\node[above][scale=1.1] at (axis cs:1.85, 1.6) {\footnotesize 1.4};
\node[above][scale=1.1] at (axis cs:2.075, 0.7) {\footnotesize 0.4};

\node[above][scale=1.1] at (axis cs:2.85, 4.9) {\footnotesize 4.7};
\node[above][scale=1.1] at (axis cs:3.075, 11.2) {\footnotesize 10.9};

\node[above][scale=1.1] at (axis cs:3.85, 1.4) {\footnotesize 1.1};
\node[above][scale=1.1] at (axis cs:4.15, 0.26) {\footnotesize 0.2};

\node[above][scale=1.1] at (axis cs:4.80, 10.5) {\footnotesize 10.1};

 \end{axis}
\end{tikzpicture}
\caption{Experimental probabilities of violation within a 95$\%$ confidence window for two (green) and three (red) orthogonal measurements per side for different criteria. The confidence window is based on the analysis performed in Sec. \ref{NSMA_biasedness}. For two (three) settings, ASR$>0.0088$ (ASR$>0.0086$) are considered to be steerable within a 95$\%$ confidence window.} 
\label{bar_plot_95_confidence_window}
\end{figure}

\subsection{E. Alternative steering inequalities} \label{other_inequalities}

In the main text, we characterised the effectiveness of the ASR inequality by comparing its probability of violation to other steering inequalities. We define those here.
\vspace{3mm}

{\it Linear steering criteria.---}
Consider that Alice and Bob are both allowed to measure $n$ observables in their sites. For unsteerable states, the following inequality should be satisfied~\cite{Saunders2010}:
\be
\left|\sum_{i=1}^n\langle A_i B_i\rangle\right| \leqslant \max_{\{a_i\}}\left[\lambda_{\max}\left(\sum_{i=1}^n a_i B_i\right)\right].
\label{Iln}
\ee
Here, $a_i = \pm 1$ and $\lambda_{\max}(X)$ is the highest eigenvalue of $X$. In this work, we consider that Alice and Bob perform $n=\{2,3\}$ observables of the form $A_i = \vec{u}_i\cdot\vec{\sigma}$ and  $B_i = \vec{v}_i\cdot\vec{\sigma}$, respectively, with $\vec{\sigma}=(\sigma_1,\sigma_2,\sigma_3)$ the vector composed of the Pauli matrices. If Bob's measurements are orthogonal, we recover the linear inequalities derived in~\cite{Cavalcanti2009}, with bound $\sqrt{n}$, for $n=\{2,3\}$.
\vspace{3mm}

{\it CHSH-like steering criteria.---}
More recently, another inequality was proposed in~\cite{Cavalcanti2015}, in a scenario in which Alice performs two dichotomic measurements while Bob performs two mutually unbiased qubit measurements. These authors then derived the following CHSH-LS inequality:
\be
\Big[\sqrt{f_+(\rho,\mu)} + \sqrt{f_-(\rho,\mu)}\Big]\leqslant 2,
\label{ICHSH}
\ee
where $f_{\pm}(\rho,\mu)=\langle (A_1\pm A_2)B_1\rangle^2 + \langle (A_1\pm A_2)B_2\rangle^2$.
\vspace{3mm}

{\it Dimension-bounded steering.---}
Bringing practical \textcolor{black}{one-sided} device independent protocols closer to fully-device independent protocols imposes serious experimental efforts. A promising candidate to overcome these limitations are DBS inequalities~\cite{Moroder2016}. Instead of trusting Bob's measurement devices, one can simply make the assumption that they act on a qubit system. Interestingly, one of the most basic steering protocols---three orthogonal qubit measurements acting on a singlet state---was shown to have an unaffected noise tolerance when decreasing the trust on Bob's side. Demonstration of steering in such scenarios is referred to as DBS. This leads to the DBS inequality (for details see~\cite{Moroder2016})
\begin{align}\label{dimbound}
    |\text{det}D|\leq\frac{1}{\sqrt{d_A}}\Big(\frac{\sqrt{2 d_A}-1}{m\sqrt{d_A}}\Big)^{m},
\end{align}
where $m$ is the number of Bob's measurements, $D$ is the measurement correlation matrix, and $d_A$ is the dimension of the chosen operators.
Interestingly, in order to derive this inequality, all assumptions on Bob's apparatus or measurements were removed, with the exception that all measurements operate in the same Hilbert space of dimension $d_{\mathcal{B}}$. 
\vspace{3mm}

{\it Rotationally invariant steering inequalities.---}
Any LHS model for this case must satisfy the steering inequality~\cite{Wollmann2018}
\be\label{Iris}
||M||_{\textrm{tr}} = \textrm{tr}\sqrt{M^T M} \leqslant \sqrt{m},
\ee
where $M_{jk} = \langle A_j B_k\rangle$ and $m$ is the number of Alice's measurements. Here, Bob's measurements are required to be composed by an orthogonal set.

\subsection{F. Adapted steering robustness - additional experimental \& simulation results}\label{additional_simulation_results}

In this section, we present further experimental results and additional simulation results. Firstly, in Table~\ref{tab:table}, we present our experimental results with and without applying the NSMA. The experimental probability distribution contains signaling (due to finite statistics) which leads to false positive outcomes in the quantum steering test, as indicated by a higher probability of violation for all considered inequalities before we apply the NSMA. Further, we note that the ASR is able to exploit the signaling to achieve 100\% probability of violation. The values in the table after the NSMA is applied are the same as those presented in Fig.~3 in the main text. 

\begin{table}[h!]
  \centering
  \begin{tabular}{|l|c | c|}
    \hline
    Inequality: & \multicolumn{2}{c|}{Probability of violation (\%):} \\
    \hline
     & With Signaling & After applying NSMA (Fig. 3) \\
    \hline
    \textbf{ASR3} & \textbf{100} & \textbf{75.5} \\
    \hline
    LS3 & 7.77 & 0.25 \\
    \hline
    DBS3 & 44.43 & 6.37 \\
    \hline
    RIS3 & 64.68 & 11.33 \\
    \hline 
    \textbf{ASR2} & \textbf{100} & \textbf{32.6} \\
    \hline
    LS2 & 6.80 & 1.14 \\
    \hline
    DBS2 & 13.77 & 2.31 \\
    \hline
    RIS2 & 25.23 & 4.99 \\
    \hline
    CHSH-LS & 48.12 & 10.73 \\
    \hline
  \end{tabular}
  \caption{Table showing the probability of violating various inequalities for 2 and 3 measurement settings. The considered inequalities are: linear steering (LS), dimension bounded steering (DBS), rotationally-invarient steering (RIS) and Clauser-Horne-Shimony-Holt-like steering (CHSH-LS). All values are calculated from the experimental data. We present both the original data (containing signaling) and the corrected signaling-free data (found by applying the NSMA to the original data).}
  \label{tab:table}
\end{table}

Next, we provide further numerical analysis of the ASR. We simulate the NS probability distribution that arises from a maximally entangled Werner state ($\mu=1$) and 10,000 quasi-uniform (QU) random sets of three orthogonal measurements on Alice's side and three Pauli measurements, ${\sigma_x,\sigma_y,\sigma_z}$, on Bob's side. Using this simulated probability distribution, we calculate the ASR and compare it to other inequalities (defined in supplementary material~\ref{other_inequalities}) for the two and three measurement settings per side cases. Considering this scenario, allows us to investigate our protocol for an ideal, noiseless scenario. We note that for two measurement settings per side, there are 90,000 possible pairs of random orthogonal settings. This is because we have 10,000 sets of three QU orthogonal settings and from this, we can select 90,000 orthogonal pairs. Due to the construction of the QU distribution, 10,000 of the pairs lie in orthogonal planes, leading to a $0\%$ probability of violation for all steering inequalities. As this is not representative of a uniform distribution, we remove these settings from the simulation. We average over the remaining 80,000 pairs of random orthogonal settings.

\begin{figure}
\centering
 \begin{tikzpicture}
 \begin{axis}[
ybar=3pt,
bar width=10.5pt,
height=6.45cm,
width=0.8\textwidth,
xmin=1.2,
xmax=4.5,
ymin=6,
ymax=100,
enlarge x limits = {abs= 0.75},
enlarge y limits = {abs = 9.0},
xtick=data,
ylabel=Probability of violation ($\%$),
y label style={at={(axis cs:-0.1,0)},anchor= 
west},
y label style={scale=1.12},
x tick label style={rotate=0},
x label style={scale=1.12},
y tick label style={/pgf/number format/1000 sep=},
xticklabels={ASR, DBS, RIS, LS, CHSH-LS},
legend entries= {2 measurement settings, 3 measurement settings},
legend style={at={(1,1)},anchor=north east}
]

\addplot  [draw = {green},
    line width = .75mm,
    draw opacity=0.95,
    fill = green,
    fill opacity=0.4,
    text opacity=1
]coordinates { (1,81.80)   (2,51.56) (3,54.10) (4, 19.35) (4.9,77.70)  };

\addplot [draw = {awesome},
    line width = .75mm,
    draw opacity=0.95,
    fill = awesome,
    fill opacity=0.34,
    text opacity=1
] coordinates { (1,100)  (2,100)  (3,100) (4,11.6) };

\node[above][scale=1.1] at (axis cs:0.85, 82) {\footnotesize 81.8};
\node[above][scale=1.1] at (axis cs:1.075, 100) {\footnotesize 100};


\node[above][scale=1.1] at (axis cs:1.85, 52) {\footnotesize 51.6};
\node[above][scale=1.1] at (axis cs:2.075, 100) {\footnotesize 100};

\node[above][scale=1.1] at (axis cs:2.85, 55) {\footnotesize 54.1};
\node[above][scale=1.1] at (axis cs:3.075, 100) {\footnotesize 100};

\node[above][scale=1.1] at (axis cs:3.85, 20) {\footnotesize 19.4};
\node[above][scale=1.1] at (axis cs:4.15, 12) {\footnotesize 11.7};

\node[above][scale=1.1] at (axis cs:4.80, 77) {\footnotesize 77.7};

 \end{axis}
\end{tikzpicture}
\caption{Theoretical probabilities of violation given a maximally entangled Werner state ($\mu=1$), for two (green) and three (red) orthogonal measurements per side for different criteria. Alice's measurement settings are chosen from a quasi-uniform random distribution on the surface of the Bloch sphere, whilst Bob's measurement settings are fixed along ${\sigma_x, \sigma_y,\sigma_z}$. For the two measurement settings per side, we excluded the case which results in $0\%$ probability of violation. See main text for details.}
\label{bar_plot_perfect_singlet}
\end{figure}
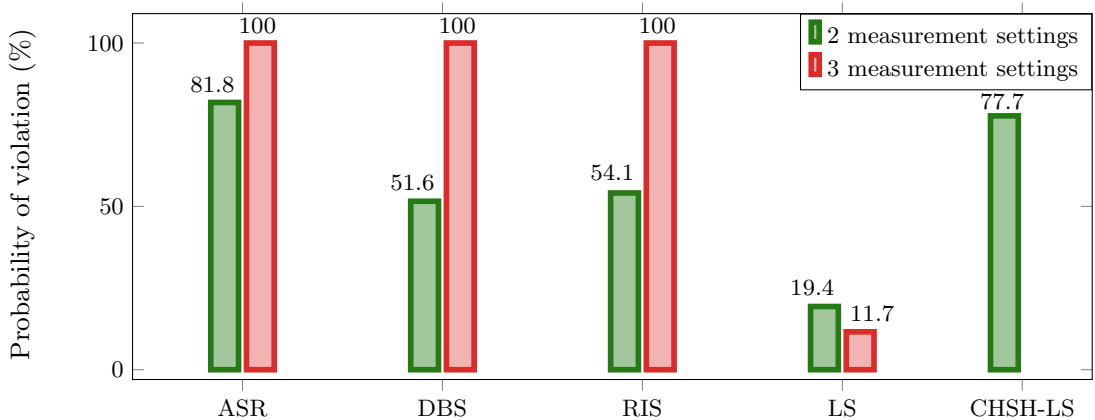

Fig.~\ref{bar_plot_perfect_singlet} compares the probability of violation for the ASR to linear-steering (LS)~\cite{Cavalcanti2009,Saunders2010}, CHSH-like steering (CHSH-LS)~\cite{Cavalcanti2015}, dimension-bounded steering (DBS)~\cite{Moroder2016} and rotationally-invariant steering (RIS). The probabilities of violating any of these inequalities are higher than in the true experiment, which is to be expected due to noise and finite statistics. It is notable that in the experiment the ASR is by far the best inequality to choose. Here, ASR-3, RIS-3 and DBS-3 all return a deterministic violation. This demonstrates the noise robustness of ASR-3.

\end{document}